\documentstyle[epsfig,aps]{revtex}
\font\eightit=cmti8                     
\def\r#1{\ignorespaces $^{#1}$}         
\begin{document}
\draft
\onecolumn
\title{
\begin{center}
  Measurement of $d\sigma/dy$ for High Mass 
  Drell-Yan $e^+e^-$ Pairs from $p\bar{p}$ Collisions at  $\sqrt{s}=1.8$~TeV
\end{center}
      }
\author{
\hfilneg
\begin{sloppypar}
\noindent
T.~Affolder,\r {21} H.~Akimoto,\r {43}
A.~Akopian,\r {36} M.~G.~Albrow,\r {10} P.~Amaral,\r 7 S.~R.~Amendolia,\r {32} 
D.~Amidei,\r {24} K.~Anikeev,\r {22} J.~Antos,\r 1 
G.~Apollinari,\r {10}
% ...
\end{sloppypar}
\begin{center}
(CDF Collaboration)
\end{center}
}
\address{
\begin{center}
\r 1  {\eightit Institute of Physics, Academia Sinica, Taipei, Taiwan 11529, 
Republic of China} \\
\r 2  {\eightit Argonne National Laboratory, Argonne, Illinois 60439} \\
\r 3  {\eightit Istituto Nazionale di Fisica Nucleare, University of Bologna,
I-40127 Bologna, Italy} \\
\r 4  {\eightit Brandeis University, Waltham, Massachusetts 02254} \\
\r 5  {\eightit University of California at Los Angeles, Los 
Angeles, California  90024} \\  
\r 6  {\eightit Instituto de Fisica de Cantabria, CSIC-University of Cantabria, 
39005 Santander, Spain} \\
\r 7  {\eightit Enrico Fermi Institute, University of Chicago, Chicago, 
Illinois 60637} \\
\r 8  {\eightit Joint Institute for Nuclear Research, RU-141980 Dubna, Russia}
\\
\r 9  {\eightit Duke University, Durham, North Carolina  27708} \\
\r {10}  {\eightit Fermi National Accelerator Laboratory, Batavia, Illinois 
60510} \\
\r {11} {\eightit University of Florida, Gainesville, Florida  32611} \\
\r {12} {\eightit Laboratori Nazionali di Frascati, Istituto Nazionale di Fisica
               Nucleare, I-00044 Frascati, Italy} \\
\r {13} {\eightit University of Geneva, CH-1211 Geneva 4, Switzerland} \\
\r {14} {\eightit Harvard University, Cambridge, Massachusetts 02138} \\
\r {15} {\eightit Hiroshima University, Higashi-Hiroshima 724, Japan} \\
\r {16} {\eightit University of Illinois, Urbana, Illinois 61801} \\
\r {17} {\eightit The Johns Hopkins University, Baltimore, Maryland 21218} \\
\r {18} {\eightit Institut f\"{u}r Experimentelle Kernphysik, 
Universit\"{a}t Karlsruhe, 76128 Karlsruhe, Germany} \\
\r {19} {\eightit Korean Hadron Collider Laboratory: Kyungpook National
University, Taegu 702-701; Seoul National University, Seoul 151-742; and
SungKyunKwan University, Suwon 440-746; Korea} \\
\r {20} {\eightit High Energy Accelerator Research Organization (KEK), Tsukuba, 
Ibaraki 305, Japan} \\
\r {21} {\eightit Ernest Orlando Lawrence Berkeley National Laboratory, 
Berkeley, California 94720} \\
\r {22} {\eightit Massachusetts Institute of Technology, Cambridge,
Massachusetts  02139} \\   
\r {23} {\eightit Institute of Particle Physics: McGill University, Montreal 
H3A 2T8; and University of Toronto, Toronto M5S 1A7; Canada} \\
\r {24} {\eightit University of Michigan, Ann Arbor, Michigan 48109} \\
\r {25} {\eightit Michigan State University, East Lansing, Michigan  48824} \\
\r {26} {\eightit University of New Mexico, Albuquerque, New Mexico 87131} \\
\r {27} {\eightit The Ohio State University, Columbus, Ohio  43210} \\
\r {28} {\eightit Osaka City University, Osaka 588, Japan} \\
\r {29} {\eightit University of Oxford, Oxford OX1 3RH, United Kingdom} \\
\r {30} {\eightit Universita di Padova, Istituto Nazionale di Fisica 
          Nucleare, Sezione di Padova, I-35131 Padova, Italy} \\
\r {31} {\eightit University of Pennsylvania, Philadelphia, 
        Pennsylvania 19104} \\   
\r {32} {\eightit Istituto Nazionale di Fisica Nucleare, University and Scuola
               Normale Superiore of Pisa, I-56100 Pisa, Italy} \\
\r {33} {\eightit University of Pittsburgh, Pittsburgh, Pennsylvania 15260} \\
\r {34} {\eightit Purdue University, West Lafayette, Indiana 47907} \\
\r {35} {\eightit University of Rochester, Rochester, New York 14627} \\
\r {36} {\eightit Rockefeller University, New York, New York 10021} \\
\r {37} {\eightit Rutgers University, Piscataway, New Jersey 08855} \\
\r {38} {\eightit Texas A\&M University, College Station, Texas 77843} \\
\r {39} {\eightit Texas Tech University, Lubbock, Texas 79409} \\
\r {40} {\eightit Istituto Nazionale di Fisica Nucleare, University of Trieste/
Udine, Italy} \\
\r {41} {\eightit University of Tsukuba, Tsukuba, Ibaraki 305, Japan} \\
\r {42} {\eightit Tufts University, Medford, Massachusetts 02155} \\
\r {43} {\eightit Waseda University, Tokyo 169, Japan} \\
\r {44} {\eightit University of Wisconsin, Madison, Wisconsin 53706} \\
\r {45} {\eightit Yale University, New Haven, Connecticut 06520} \\
\end{center}
}
\maketitle

\begin{abstract}
We report on the first measurement of the rapidity distribution 
$d\sigma/dy$ over nearly the entire kinematic region of rapidity
for $e^+e^-$ pairs in 
the $Z$-boson region of $66<M_{ee}<116$~GeV$/c^2$ and at higher 
mass  $M_{ee}>116$~GeV$/c^2$. The data sample consists of
108~pb$^{-1}$ of $p\bar{p}$ collisions at $\sqrt{s}=1.8$~TeV
taken by the Collider Detector at Fermilab during 1992--1995.  
The total cross section in the $Z$-boson region is measured to be 
$252 \pm 11$ pb.
The measured total cross section and  $d\sigma/dy$ are compared with 
quantum chromodynamics calculations in leading and higher orders.
\end{abstract}
\pacs{PACS numbers: 13.85.Qk, 12.38.Qk}
\twocolumn
%
%
\par
Most measurements at high energy proton-antiproton colliders are 
performed in the central rapidity production region, $|y| < 1$. 
A model dependent extrapolation for $|y|>1$ is needed to
extract the total cross section for hard processes such as
top quark production or $W$ and $Z$ boson production.
This extrapolation is made using Monte Carlo programs
(e.g. {\small PYTHIA}~\cite{Pythia6104}), which incorporate
quantum chromodynamics (QCD) calculations in leading order (LO)
or next to leading order (NLO). A previous measurement
of the rapidity distribution, $d\sigma/dy$, for dimuon pairs 
in the $Z$-bosons mass region was limited to $|y| < 1$~\cite{DiMuPRD}.
In this communication, we present the first measurement of 
$d\sigma/dy$ for $e^+e^-$ pairs in the $Z$-boson mass and high mass
region over nearly the entire kinematic region of rapidity. 
At the Tevatron $p\bar p$ collider, the kinematic limit at the
$Z$-boson mass is $|y|=3.0$, while we measure $|y|$ up to 2.8.
The $d\sigma/dy$ distributions are compared to the predictions of 
QCD in LO and NLO. This measurement is also relevant for precision 
$W$ boson mass measurements at hadron colliders, where $W$'s are 
reconstructed using $e\nu$ and $\mu\nu$ pairs from the Drell-Yan process. 
\par
In hadron-hadron collisions at high energies, massive $e^+e^-$ 
pairs are produced via the Drell-Yan~\cite{DrellYan} process. 
In the standard model, quark-antiquark annihilation form an 
intermediate $\gamma^*$ or $Z$ ($\gamma^*/Z$) vector boson, 
which then decays into an $e^+e^-$ pair. In LO, 
the momentum fraction $x_1$ ($x_2$) of the partons in the 
proton (antiproton) are related to the rapidity~\cite{CdfDet}, 
$y$, of the boson via the equation $ x_{1,2} = ({M}/{\sqrt{s}}) e^{\pm y}$.
Here $s$ is the center of mass energy, and $M$ is the mass of the 
dilepton pair. Therefore, dilepton pairs which are produced at large
rapidity originate from events in which one parton is
at large $x$ and another parton is at very small $x$.
Since the quark distributions for $x$ up to 0.9 are well constrained by the 
deep-inelastic lepton scattering experiments~\cite{dupaper},
comparisons of data and theory for $d\sigma/dy$, and the total
cross sections provide a test of the theory, \mbox{e.g.} missing 
NNLO~\cite{NNLO,Bodek} or power correction~\cite{power} terms.
\par
The $e^+e^-$ pairs are from 108~pb$^{-1}$ of $p\bar{p}$ collisions
at $\sqrt{s}=1.8$~TeV taken by the Collider Detector at Fermilab~\cite{CdfDet}
(CDF) during 1992--1993 ($18.7 \pm 0.7$~pb$^{-1}$) and 1994--1995 
($89.1 \pm 3.7$~pb$^{-1}$). CDF is a solenoidal magnetic spectrometer 
surrounded by projective-tower-geometry calorimeters and outer muon 
detectors. Only detector components used in this measurement are described 
here. Charged particle momenta and directions are measured by the 
spectrometer, which consists of a 1.4~T axial magnetic field, an 84-layer
cylindrical drift chamber (CTC), an inner vertex tracking chamber (VTX),
and a silicon vertex detector (SVX). The polar coverage of the CTC 
tracking is $|\eta|<1.2$. The $p\bar{p}$ collision point along the beam 
line $(Z_{\rm vertex})$ is determined using tracks in the VTX. 
The energies and directions~\cite{CdfDet} of electrons, photons, 
and jets are measured by three separate calorimeters covering three regions: 
central $(|\eta|<1.1)$, end plug $(1.1<|\eta|<2.4)$, and forward 
$(2.2<|\eta|<4.2)$. Each region has an electromagnetic (EM) and
hadronic (HAD) calorimeter. 
\par
In a previous letter, we presented $d\sigma/dP_{\rm T}$ of $Z$
boson~\cite{willis}. This analysis is an extension of the
$d\sigma/dP_{\rm T}$ measurement. The
$d\sigma/dP_{\rm T}$ analysis has three categories of $e^+e^-$ pairs:
central-central (CC), central-end plug (CP), and central-forward (CF).
This analysis extends the sample to the forward rapidity region by 
including plug-plug (PP) and plug-forward (PF) events. The
inclusion of these events increases the event sample by $20\%$ and
allows for measurement of $Z$ bosons with $|y|$ up to 2.8.
An improvement in this analysis is the additional VTX tracking
requirements for plug and forward electrons.
The VTX covers the entire rapidity range in this study, and plays
an important role in removing background in the high $\eta$ region
which is not covered by the CTC.
\par
The sample of $e^+e^-$ events was collected by a three-level online 
trigger that required an electron in either the central or the plug 
calorimeter. The offline analysis selects events with two or more 
electron candidates.
Since the electrons from the Drell-Yan process are typically isolated, both
electrons are required to be isolated from any other activity in the
calorimeters. Electrons in the central, end plug, and forward regions 
are required to be within the fiducial area
of the calorimeters and have a minimum
$E_{\rm T}$ of 22, 20, and 15~GeV, respectively. 
To improve the purity of the sample, electron identification cuts
are applied~\cite{willis}. 
For CC, CP, and CF events, the central electron (or one of them if 
there are two) is required to pass strict criteria. The criteria 
on the other electron are looser. A central electron must have a 
CTC track that extrapolates to the electron's shower clusters in 
the EM calorimeter. These clusters must have EM-like transverse 
shower profiles. The track momentum and the EM shower energy must 
be consistent with one another. The track is also used to determine 
the position and direction of the central electron. The fraction of 
energy in the HAD calorimeter towers behind the EM shower is required 
to be consistent with that expected for an EM shower 
($E_{\rm HAD}/E_{\rm EM}$). 
The plug electrons must also have an EM-like transverse 
shower profile. The end plug and forward electrons 
are required to pass the $E_{\rm HAD}/E_{\rm EM}$ requirement and to
have a track in the VTX which originates from the same vertex as the 
other electron and points to the position of the electromagnetic cluster 
in the calorimeter. The ratio of found to expected hits in the VTX
is required to be greater than $70\%$ and $50\%$ for plug and 
forward electrons, respectively. The VTX tracking efficiency is 
$(97.8\pm 0.3)\%$ for plug electrons and $(97.0\pm 0.9)\%$ for 
forward electrons. This requirement on plug and forward
electrons reduce the background rates from $11\%$ to $2\%$ for 
PP events, and from $22\%$ to $4\%$ for PF events. 
\par
The data sample is divided into two mass bins: 
the $Z$ region ($66<M_{ee}<116$~GeV/$c^2$) and the high mass
region ($M_{ee}>116$~GeV/$c^2$). After all cuts, the numbers of CC, CP, 
CF, PP, and PF events in the $Z$ mass region are 2894, 3811, 621, 
1236, and 589, respectively. The backgrounds are low and are 
estimated using the data. The backgrounds in the CP, CF, PP, and PF
topologies are dominated by jets and are $28 \pm 5$, $13 \pm 3$, 
$24 \pm 2$, and $23 \pm 3$ events, respectively. Because of the CTC tracking 
requirement, the jet background for the CC sample is negligible. 
The CC background is mainly from $e^+e^-$ pairs from $W^+W^-$, 
$\tau^+\tau^-$, $c\bar{c}$, $b\bar{b}$, and $t\bar{t}$ sources. 
This background, estimated using $e^\pm\mu^\mp$ pairs~\cite{EMuSel}, 
is $3 \pm 2$ events. In the high mass bin, the numbers of CC, CP, 
CF, PP, and PF events are 61, 59, 9, 18, and 5, respectively. 
The mean mass of the high mass bin is 152.6 ~GeV/$c^2$.
\par
The acceptance for Drell-Yan
$e^+e^-$ pairs is obtained using the Monte Carlo event generator,
{\small PYTHIA}~\cite{Pythia6104}, and CDF detector simulation programs.
{\small PYTHIA} generates the LO QCD
interaction ($q+\bar{q} \rightarrow \gamma^*/Z$), simulates initial state QCD
radiation via its parton shower algorithms, and generates the decay,
$\gamma^*/Z \rightarrow e^+e^-$. To approximate higher order QCD corrections
to the LO mass distribution, a ``$K$-factor''~\cite{AEM-Kfactor}
is used as an event weight:
$K(M^2) = 1 + \frac{4}{3}(1 + \frac{4}{3}\pi^2) \alpha_s(M^2)/2\pi$,
where $\alpha_s$ is the two loop QCD coupling. This factor improves the
agreement between the NLO and LO mass spectra.
(For $M>50$~GeV/$c^2$, $1.25 < K < 1.36$.)
The CTEQ3L~\cite{Cteq3pdf} PDFs are used in the acceptance calculations. 
Final state QED 
radiation~\cite{QEDRadCor} from the $\gamma^*/Z \rightarrow e^+e^-$ vertex is
added by the {\small PHOTOS}~\cite{Photos20} Monte Carlo program. 
Generated events are
processed by CDF detector simulation programs and are reconstructed as data.
The calorimetry energy scales and resolutions used in the detector
simulation are extracted from the data, as are the cut efficiencies and
corresponding errors. Simulated events are accepted if after the reconstruction
they pass the $e^+e^-$ pair mass, the detector fiducial, 
and kinematic ($E_T$) cuts.
\par
\begin{figure}
\begin{center}
\mbox{\epsfxsize=3.3in \epsffile{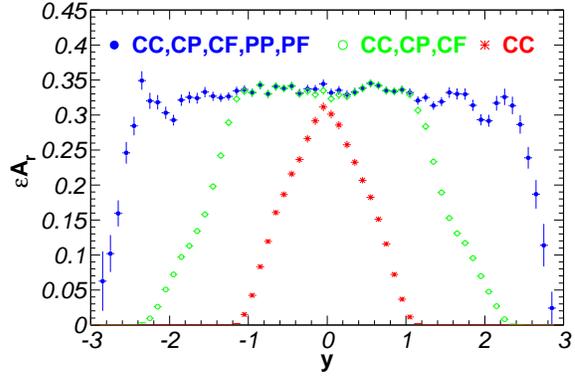}}   
\end{center}
\caption{The efficiency times acceptance for $e^+e^-$ pairs
in the $Z$ boson mass region: (a) for the CC sample (asterisks), (b)
for the CC+CP+CF sample (open circles),
and (c) for the CC+CP+CF+PP+PF sample (solid circles).}
\label{fig1}
\end{figure}

In the $d\sigma/dy$ measurement, various samples are  
combined and binned in $y$. The $d\sigma/dy$ is calculated with
\begin{displaymath}
 \frac{d\sigma}{dy} =
 \frac{\Delta N} {C \: \Delta y  \:
                            \sum_r {\cal L}_r \: \epsilon A_r} .
\end{displaymath}
The $\Delta N$ is the background-subtracted event count in a $y$ bin,
$C$ is a bin centering correction, $\Delta y$ is the bin width, the sum
$r$ is over the 1992--1993 and 1994--1995 runs, ${\cal L}_r$ is the 
integrated luminosity, and $\epsilon A_r$ is the run's combined event selection
efficiency and acceptance. The backgrounds subtracted from the event count are
predicted using the data and background samples. The factor $C$ corrects the
average value of the cross section in the bin to its bin center value.
Acceptances are calculated separately for CC, CP, CF, PP, and PF pairs.
They are combined with the corresponding event selection efficiencies 
to give $\epsilon A_r$. Figure~\ref{fig1} shows the $\epsilon A_r$ for
events in the $Z$ region as a function of $y$ for the CC sample, 
the CC+CP+CF sample, and the CC+CP+CF+PP+PF sample, respectively. 
It indicates that the PP+PF events extend the acceptance of the $y$ 
measurement to $|y|=2.8$. We significantly increase our statistics by 
extending the acceptance beyond $y > 1.2$.
The $d\sigma/dy$ of $e^+e^-$ events in the $Z$ mass region is shown 
in Table~\ref{Table1}.
\par
The systematic errors considered are from variations in the background
estimates using different methods, the background in the efficiency
sample, the uncertainty in energy resolution of the calorimeter, 
the choice of PDFs, and the distribution of $Z$ $P_{\rm T}$ used in the
Monte Carlo event generator.  The systematic error
from the calorimeter resolution is $0.2\%$ in the low $y$ region
and increases to $0.7\%$ at $|y|>2.5$. The systematic error from 
variations in the $Z$ $P_{\rm T}$ is $0.5\%$ at $|y|=0$ and is $2.0\%$ 
at $|y|=2.8$. The systematic error from the choice of PDFs is less 
than $1\%$ in $|y|<2.0$ and increases to $2\%$ at $|y|=2.8$. 
For the total cross section measurement,
the combined systematic error is $0.6\%$ (excluding the luminosity
uncertainty). The $p\bar{p}$ collision luminosity is derived
with CDF's beam-beam
cross section, $\sigma_{\rm BBC}=51.15 \pm 1.60$~mb~\cite{CDFLum,CDFLum1}.
The luminosity error of 3.9$\%$ contains the $\sigma_{\rm BBC}$ 
error and uncertainties specific to running conditions.
\par
\begin{figure}
\begin{center}
\mbox{\epsfxsize=3.37in \epsfysize=2.25in \epsffile{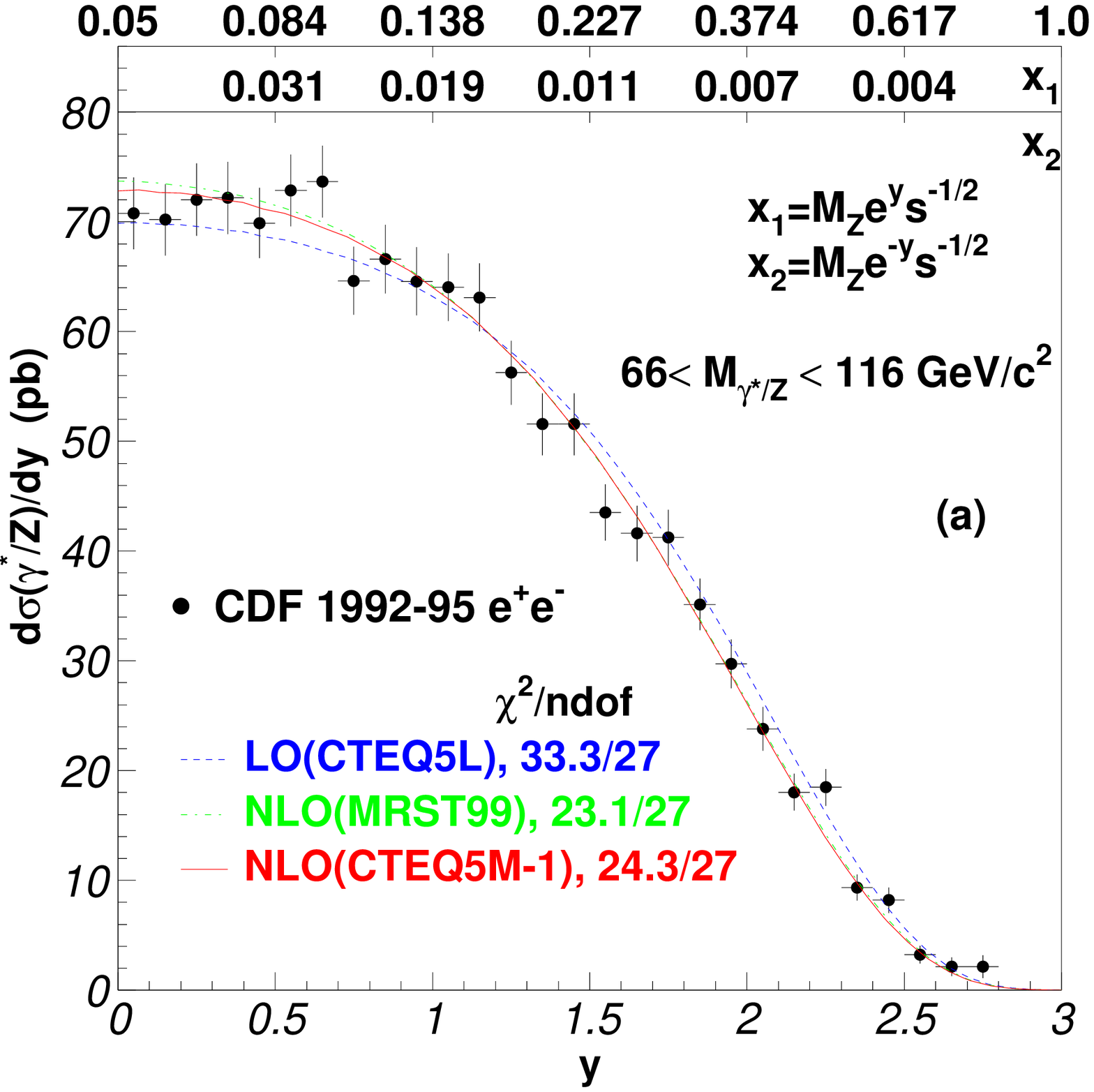}}   
\mbox{\epsfxsize=3.37in \epsfysize=2.25in \epsffile{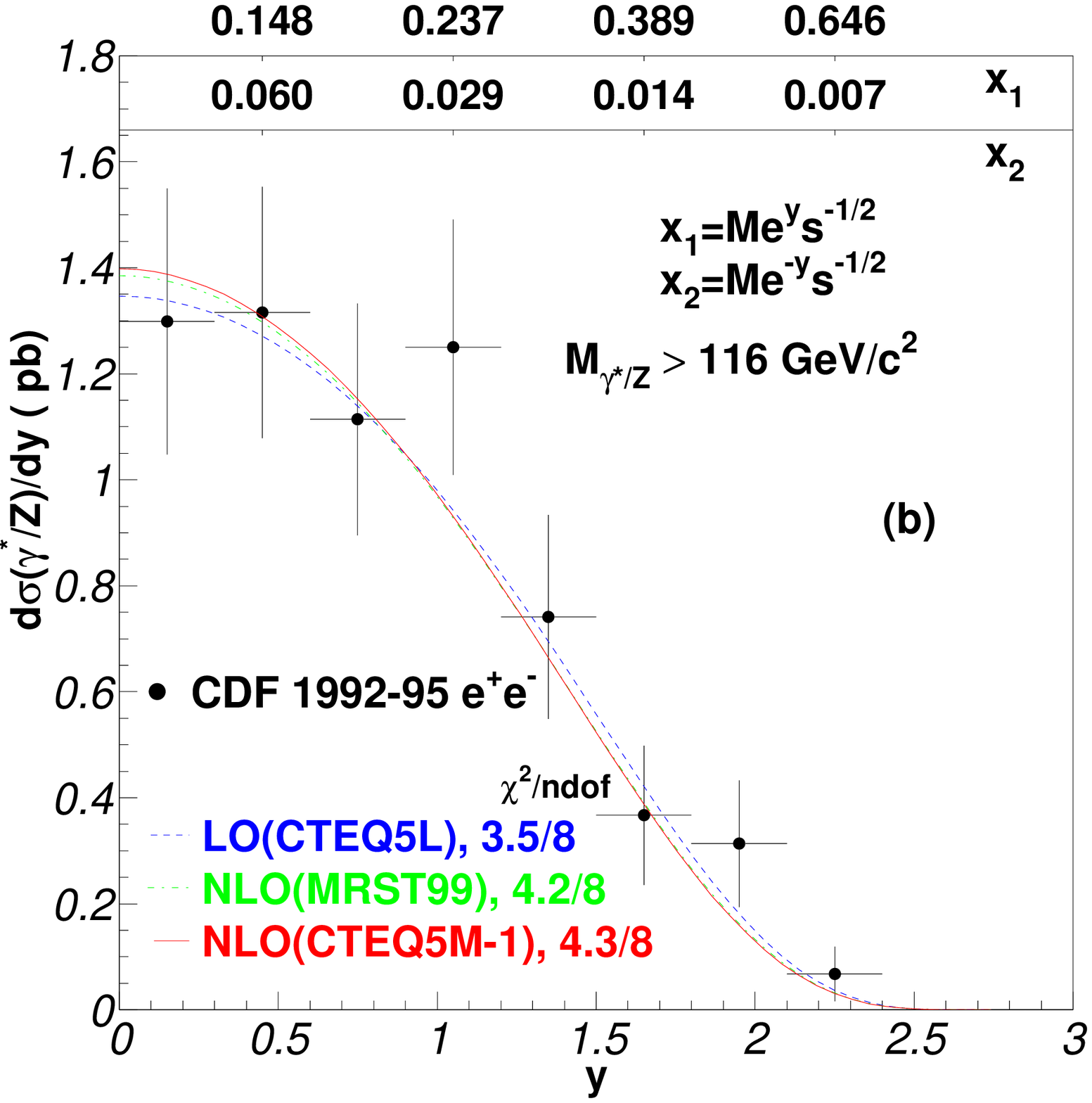}} 
\end{center}
\caption{$d\sigma/dy$ 
distribution of $e^+e^-$ pairs: (a) in the $Z$ boson mass 
($66<M_{ee}<116$~GeV/$c^2$) region. 
(b) in the high mass ($M_{ee}>116$~GeV/$c^2$) region. The $M$ used to
obtain $x_1$ and $x_2$ in (b) is the mean mass over the bin. The error
bars on the data include statistical errors only. The
theoretical predictions have been normalized to the data in the
$Z$ boson mass region.}
\label{fig2}
\end{figure}

Figures~\ref{fig2}(a) and \ref{fig2}(b) compare the measured 
$d\sigma/dy$'s to theoretical predictions in the $Z$ mass 
and high mass regions, respectively. The top horizontal axes on
these figures are the corresponding values of the $x_1$ and $x_2$ 
as a function of $y$. The predictions are LO calculations with 
CTEQ5L~\cite{CTEQ5} PDFs and NLO~\cite{NLO} calculations with MRST99~\cite{MRS} 
and CTEQ5M-1~\cite{CTEQ5} PDFs.
The predictions in Figure~\ref{fig2}(a) have been normalized by a 
factor ``F'', the ratio of measured total cross section to 
the prediction (F=1.51, 1.14, and 1.13 for the CTEQ5L,
MRST99 and CTEQ5M-1 PDFs, respectively).
The predictions in Figure~\ref{fig2}(b) are normalized to the 
data using the factor F from the $Z$ mass region. 
We compare the data to the theory using statistical errors only.
As the $\chi^2$ values listed in Figure~\ref{fig2}(a) indicate,
the LO calculation using recent LO PDFs does not fit the shape
as well as the NLO calculation with the most recent NLO PDFs.
\par
Model independent measurements of the total production cross sections 
for  $e^+e^-$ pairs are extracted by integrating the measured values 
of $d\sigma/dy$. Because there are no data for $|y|>2.8$, we use
a NLO calculation with the CTEQ5M-1 PDFs to estimate $\gamma^*/Z$ 
production in that region. The cross section in the region of 
$|y|>2.8$ is about $0.02\%$ of the predicted total cross section.
The extracted cross section in the $Z$ mass region is 
$252.1 \pm 3.9\:{\rm (stat.)} \pm 1.6\:{\rm (syst.)} \pm 9.8\:{\rm (lum.)}$~pb.
The corresponding $\sigma(p\bar p\rightarrow Z) \cdot Br(Z\rightarrow
ee)$ is $253 \pm 4({\rm stat.+syst.}) \pm 10 {\rm (lum.)}$~pb.
This measurement is in agreement with our previous measurements
in the dielectron~\cite{willis} 
($248 \pm 5 {\rm (stat.+syst.)} \pm 10 {\rm (lum.)}$~pb,
using only the CC+CP+CF $e^+e^-$ sample), and dimuon~~\cite{DiMuPRD} 
($237 \pm 9 {\rm (stat.+syst.)} \pm 9{\rm (lum.)}$~pb using a CC sample)
channels. These previous measurements use a QCD model to correct
for the missing events at high rapidity.
The combined $e^+e^-$ and $\mu^+\mu^-$ cross section is 
$250 \pm 4$(stat.+syst.)$\pm 10$ pb.
Since the $p\bar{p}$ inelastic cross section used by CDF in luminosity
calculations differs from \mbox{DO\hspace*{-1.5ex}/}{\hspace*{0.5ex}'s} by 
$+5.9\%$~\cite{CDFLum1},
measured cross sections must be renormalized before comparisons to 
other experiments. The \mbox{DO\hspace*{-1.5ex}/}
measurement~\cite{D0WZ1b} of $221 \pm 11$~pb when renormalized to
the CDF luminosity measurement is $234 \pm 12$~pb.
\par
The total cross section measurements can also be compared to QCD calculations.
Fixed order QCD calculations have uncertainties from PDF measurements and
corrections from higher orders of perturbation theory,
\mbox{i.e.}, the $K$-factor.
The NLO-to-LO total cross section correction is significant: $K \sim 1.4$.
In contrast, the NLO total cross section is lower than NNLO~\cite{NNLO}
prediction by only $2.3\%$. The NNLO prediction
with the latest NLO MRST99 PDFs is $227 \pm 9$~pb,
where the $4\%$ error is mostly from uncertainties~\cite{MRS} in the NLO PDFs.
Although a full set of NNLO PDFs is not available, recent 
estimates~\cite{Bodek}
of NNLO PDFs indicate that the NNLO PDFs will increase the theoretical
cross sections by $5\%$.
Given these uncertainties, the theoretical expectation is consistent with
the $Z$ cross section measurements.
\par
The measurement of the Drell-Yan total cross section in the high mass
region is $4.0 \pm 0.4\:{\rm (stat.+syst.)} \pm 0.2\:{\rm (lum.)}$~pb.
The corresponding prediction of the total cross section from the NNLO QCD 
theory using MRST99 PDFs is $3.3$ pb. 
\par
In summary, the rapidity distributions of $e^+e^-$ pairs in the $Z$ boson
mass and high mass region have been measured for the 
first time over nearly the entire kinematic region.
This measurement uses a new tracking technique in the high 
rapidity region to reduce the background and uncertainties associated with it. 
In addition, unlike the previous measurement of the total cross section, this
measurement is model independent.
\par
The vital contributions of the Fermilab staff and the technical staffs of the
participating institutions are gratefully acknowledged. This work is supported
by the U.S. Department of Energy and National Science Foundation;
the Natural Sciences and Engineering Research Council of Canada;
the Istituto Nazionale di Fisica Nucleare of Italy;
the Ministry of Education, Science and Culture of Japan;
the National Science Council of the Republic of China;
the A.P. Sloan Foundation, and the Swiss National Science Foundation.
%

%
%
\begin{table}
\caption{$d\sigma/dy$ distribution of $e^+e^-$ events in the mass 
range $66<M_{ee}<116$~GeV/$c^2$. The first and second
errors are statistical and systematic, respectively.
The $3.9\%$ luminosity error is not included.
Here $y$ is the bin center value.}
\label{Table1}
\begin{center}
\begin{tabular}{crcr}
\multicolumn{1}{c}{$y$} & \multicolumn{1}{c}{$d\sigma/dy$ [pb]} &
\multicolumn{1}{c}{$y$} & \multicolumn{1}{c}{$d\sigma/dy$ [pb]} \\ \hline
 0.05  & $ 70.78 \pm 3.27 \pm 0.37$  & 1.45  & $ 51.56 \pm 2.82 \pm 0.45$  \\
 0.15  & $ 70.19 \pm 3.26 \pm 0.37$  & 1.55  & $ 43.51 \pm 2.56 \pm 0.39$  \\
 0.25  & $ 72.03 \pm 3.30 \pm 0.41$  & 1.65  & $ 41.62 \pm 2.54 \pm 0.45$  \\
 0.35  & $ 72.18 \pm 3.30 \pm 0.45$  & 1.75  & $ 41.24 \pm 2.53 \pm 0.46$  \\
 0.45  & $ 69.90 \pm 3.20 \pm 0.50$  & 1.85  & $ 35.15 \pm 2.35 \pm 0.40$  \\
 0.55  & $ 72.86 \pm 3.25 \pm 0.59$  & 1.95  & $ 29.72 \pm 2.23 \pm 0.36$  \\
 0.65  & $ 73.68 \pm 3.27 \pm 0.64$  & 2.05  & $ 23.80 \pm 1.98 \pm 0.33$  \\
 0.75  & $ 64.64 \pm 3.09 \pm 0.58$  & 2.15  & $ 18.04 \pm 1.68 \pm 0.29$  \\
 0.85  & $ 66.59 \pm 3.13 \pm 0.63$  & 2.25  & $ 18.47 \pm 1.70 \pm 0.33$  \\
 0.95  & $ 64.59 \pm 3.10 \pm 0.67$  & 2.35  & $  9.35 \pm 1.17 \pm 0.19$  \\
 1.05  & $ 64.04 \pm 3.09 \pm 0.70$  & 2.45  & $  8.20 \pm 1.17 \pm 0.19$  \\
 1.15  & $ 63.11 \pm 3.10 \pm 0.67$  & 2.55  & $  3.24 \pm 0.80 \pm 0.08$  \\
 1.25  & $ 56.26 \pm 2.92 \pm 0.55$  & 2.65  & $  2.15 \pm 0.85 \pm 0.06$  \\
 1.35  & $ 51.57 \pm 2.82 \pm 0.45$  & 2.75  & $  2.14 \pm 1.02 \pm 0.06$  \\
\end{tabular}
\end{center}
\end{table}

\begin{references}
\bibitem{Pythia6104}
 T. Sj\"{o}strand, Comput. Phys. Commun. {\bf 82}, 74 (1994).
 The default {\small PYTHIA~6.104} is used except
 {\tt MSTJ(41)=1}, {\tt MSTP(91)=2}, and {\tt PARP(92,93)=1.25,10.0}~GeV.
\bibitem{DiMuPRD}
 F. Abe {\it et al.}, Phys. Rev. D {\bf 59}, 052002 (1999).
\bibitem{DrellYan}
 S.D. Drell and T.-M. Yan, Phys. Rev. Lett. {\bf 25}, 316 (1970).
\bibitem{CdfDet}
 F. Abe {\it et al.}, Nucl. Instrum. Methods Phys. Res. Sect. A {\bf 271},
 387 (1988). CDF coordinates are in $(\theta,\phi,z)$, where $\theta$ is the
 polar angle relative to the proton beam (the $+z$ axis), and $\phi$ the
 azimuth. The pseudorapidity is $\eta=-\ln \tan(\theta / 2)$. Here
 $P_{\rm T}=P \sin{\theta}$, 
 $y=\frac{1}{2}ln\frac{P+P_z}{P-P_z}$, $P$ and $P_z$ are the magnitude 
 and $z$ component of a particle's momentum, and
 $E_{\rm T}=E \sin{\theta}$, where $E$ is the energy measured in the
 calorimeter.
\bibitem{dupaper}
U. K. Yang and A. Bodek, Phys. Rev. Lett. {\bf 82}, 2467 (1999).
\bibitem{NNLO}
${\tt \overline{MS}}$: R. Hamberg, W.L. van Neerven, and T. Matsuura,
Nucl. Phys. B {\bf 359}, 343 (1991).
DIS :W. L. Van Neerven and E. B. Zijlstra, Nucl. Phys. B {\bf 382}, 11 (1992).
%
\bibitem{Bodek}
U.K. Yang and A. Bodek, Euro. Phys. Jour. C {\bf 13}, 241 (2000);  
W.L. van Neerven and A. Vogt, Nucl. Phys. B {\bf 568}, 263 (2000), 
hep-ph/9907472.
%
\bibitem{power}
M. Dasgupta, J. High Energy Phys. 12(1999) 008.
%
\bibitem{willis}
T. Affolder {\it et al.}, Phys. Rev. Lett. {\bf 84}, 845 (2000).
%
\bibitem{EMuSel}
 The CDF top-quark high-$P_{\rm T}$ dilepton selection 
 (F. Abe {\it et al.}, Phys. Rev. D {\bf 50}, 2966 (1994))
 is used, but with both leptons isolated and no jet cuts.
%
\bibitem{AEM-Kfactor}
 G. Altarelli, R. E. Ellis, and G. Martinelli, Nucl. Phys. B {\bf 157},
 461 (1979).
%
\bibitem{Cteq3pdf}
 H. L. Lai {\it et al.}, Phys. Rev. D {\bf 51}, 4763 (1995).
The old PDF CTEQ3L is used in the acceptance
calculation since it happens to describe the shape of
the $y$ distribution well ($\chi^2$ = 21.2/ndof, F=1.55).
%
\bibitem{QEDRadCor}
 U. Baur, S. Keller, and W. K. Sakumoto, Phys. Rev. D {\bf 57}, 199 (1998).
%
\bibitem{Photos20}
 E. Barberio and Z. Was, Comput. Phys. Commun. {\bf 79}, 291 (1994);
 E. Barberio, B. van Eijk, and Z. Was, {\it ibid.} {\bf 66}, 115 (1991).
%
\bibitem{CDFLum}
 F. Abe {\it et al.}, Phys. Rev. Lett. {\bf 76}, 3070 (1996).
%
\bibitem{CDFLum1}
D. Cronin-Hennessy {\it et al.}, Nucl. Instrum. Methods Phys. Res., Sect. A 
{\bf 443/1}, 37-50 (2000)
%
\bibitem{CTEQ5}  H. L. Lai {\it et al.}, Eur. Phys. J. C12, 375 (2000)
%
\bibitem{NLO}
P. J. Rijken and W. L. van Neerven, Phys. Rev. D. {\bf 51}, 44 (1995)
%
\bibitem{MRS} A.D. Martin  {\it et al.}, Eur. Phys. J. C14,133 (2000).
Note that the QCD evolution code used by the MRST99 and CTEQ5M-1 PDFs are
now in agreement with each other.
%
\bibitem{D0WZ1b}
 B. Abbott {\it et al.}, Phys. Rev. D. {\bf 61}, 072001 (2000).
\end{references}
\end{document}